\documentstyle[aps,multicol,prl,epsf,citesort]{revtex}

\def \BE {\begin{equation}}
\def \EE {\end{equation}}
\def \BEA {\begin{eqnarray}}
\def \EEA {\end{eqnarray}}

\def \ta {\widetilde{a}}

\begin{document}

\title{ Renormalized waves and discrete breathers in $\beta$-FPU chains}
\author{Boris Gershgorin$^1$, Yuri V. Lvov$^1$ and David Cai$^2$ \\ \ \\
$^1$ Department of Mathematical Sciences, Rensselaer Polytechnic
Institute, Troy, NY 12180\\
$^2$ Courant Institute of Mathematical Sciences,
        New York University, New York, NY 10012\\
} \maketitle
\begin{abstract}
We demonstrate via numerical simulation that in the \textit{strongly}
nonlinear limit, the $\beta$-FPU system in thermal equilibrium
behaves surprisingly like weakly nonlinear waves in properly
renormalized normal variables. This arises because the collective
effect of strongly nonlinear interactions effectively renormalizes
linear dispersion frequency and leads to effectively weak
interaction among these renormalized waves.  Furthermore, we show
that the dynamical scenario for thermalized $\beta$-FPU chains is
spatially highly localized discrete breathers riding chaotically on
spatially extended, renormalized waves.
\end{abstract}

\begin{multicols}{1}
\vskip -0.5in

The Fermi-Pasta-Ulam (FPU) lattice was introduced in their classical
work~\cite{FPU} to address fundamental issues of statistical physics
such as equipartition of energy, ergodicity.  The attempt to resolve
the mystery of the FPU recurrence (the system did not thermalize as
was expected but rather kept returning to the initial
state~\cite{FPU}) has spurred many great mathematical and physical
discoveries, such as the celebrated KAM theorem and soliton
physics~\cite{Appl}. Despite this remarkable progress, there are
still fundamental open questions that are under vigorous
debate~\cite{debate}, such as what is the route to thermalization
and how to fully characterize thermalized $\beta$-FPU system.
Furthermore, in the last decade discrete breathers (DB) as spatially
localized, time periodic lattice excitations were
discovered~\cite{Flach}. Arising from energy localization in
nonlinear lattices, they play important roles in many dynamics in
fiber optics, condensed matter physics and molecular
biology~\cite{Flach2}. The existence of DBs has been addressed
rigorously~\cite{VarProof}. 
Important conceptual issues
naturally arise, such as what is the role of DBs on the route to
equilibrium~\cite{MaxLyap} 
and how do they
manifest in thermalization of the FPU system? Resolution of these
issues will certainly provide deep insight into the fundamental
understanding of route to thermalization for general nonlinear
physical systems. Most of the results regarding DBs in $\beta$-FPU
chains have so far only addressed their behavior in the transient
state of \textit{weakly} nonlinear regimes before thermalization
occurs~\cite{PhysD,SlowRelax}.

In this Letter, we investigate the FPU dynamics in the strongly
nonlinear limit. We demonstrate that, quite surprisingly, even for
strong nonlinearity, the $\beta$-FPU system in thermal equilibrium
behaves like weakly nonlinear waves in properly chosen variables.
Such behavior results from the collective effect of strongly
nonlinear interactions effectively renormalizing linear dispersion
relation. This observation enables us to use a well-developed weak
turbulence (WT) formalism~\cite{ZLF} for the description of the
$\beta$-FPU chains even in a strongly nonlinear regime. Furthermore,
in addition to the nonlinear waves, we observe the DB excitations in
the thermalized state of $\beta$-FPU chains. Previously such DBs
were observed only during transient stages towards
thermalization~\cite{PhysD}. Here we show via numerical simulation
that DBs actually persist and coexist with renormalized waves
 in the thermalized state. Thus, in the thermalized $\beta$-FPU, there are two
kinds of quasi-particle excitations, one localized in $k$-space as
renormalized nonlinear waves/phonons, and the other, localized in
$x$-space as DBs.

The $\beta$-FPU chain is described by the Hamiltonian,
\BEA
H&=&H_2+H_4,\label{Ham_pq}\\
H_2&=&\frac{1}{2}\sum_{i=1}^N{p_i^2}+(q_i-q_{i+1})^2;~~H_4=\frac{\beta}{4}\sum_{i=1}^N(q_i-q_{i+1})^4,\nonumber
\EEA
with periodic boundary conditions,
where $\beta$ is \textit{a nonlinear parameter}, $p_i$ and $q_i$,
are the $i$-th particle momentum and displacement from the
equilibrium position, respectively. In terms of $P_k$ and $Q_k$, the Fourier transforms of
$p_i$ and $q_i$, the Hamiltonian becomes
\BEA
H=\frac{1}{2}\sum_{k=1}^{N-1}\left[|P_k|^2+\omega_{k}^2|Q_k|^2\right]+V(Q),\label{Ham_PQ}
\EEA
where $\omega_k=2\sin\frac{\pi k}{N}$ is the linear dispersion
frequency and $V(Q)$ is the Fourier transform of $H_4$.  $V(Q)$ is a
linear combination of various quartic products of $Q_k$'s and
$Q_k^*$'s. We numerically integrate the canonical equations of
motion for Hamiltonian (\ref{Ham_pq}) to study possible
dynamical scenario of the FPU~\cite{NumInt}. Since the thermalized
state is our primary interest we chose random initial data~\cite{NumInt}.
 Note that the behavior of the $\beta$-FPU in
equilibrium is fully characterized by only one parameter $\beta
H$~\cite{ExFPU} where $H$ is the total energy of the system. We
fixed $H=200$ and $N=128$ (except for Fig.~\ref{fig7_eqp}, where
N=1024) and varied $\beta$.

\begin{figure}
\epsfxsize=2.3in \epsfysize=2.in \epsffile{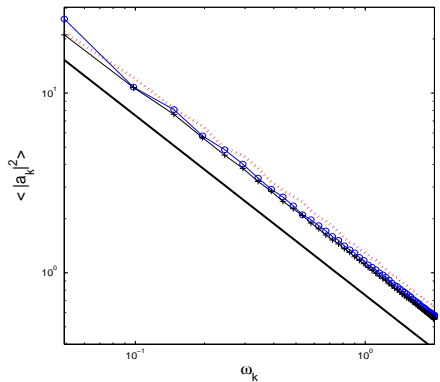}
 \vskip 0.1in
\caption[]{\footnotesize\baselineskip11pt Power spectra for the
thermalized state for relatively strong nonlinearity
$\beta=1$(dots), $\beta=8$(circles) and $\beta=32$(pluses). The
thick solid line is $1/\omega_k$ for comparison. Note that
time window $T = 10^5$  was used for averaging the power
spectrum.
} \label{fig1_ak}
\end{figure}

\begin{figure}
\vskip -0.3in \epsfxsize=2.7in \epsfysize=2.in
 \epsffile{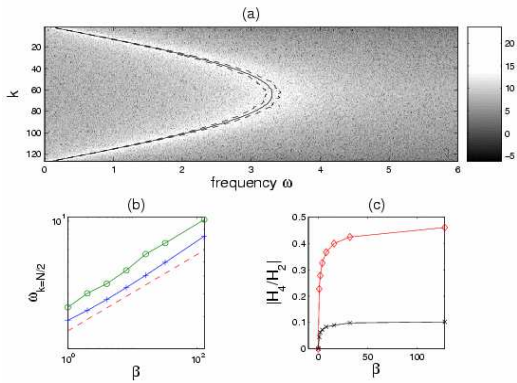}
 \vskip 0.1in
\caption[]{\footnotesize\baselineskip11pt (a) Density plot of
spectrum for $\beta=1$ and $H=200$ ($\ln(|a_k(\omega)|^2)$ is
plotted). (b) renormalized linear frequency as a function of
$\beta$: analytical predictions (pluses) and numerical measurements
(circles). $\beta^{0.2}$ is shown for comparison (dashed). (c)
effective nonlinearity as a function of $\beta$ before (diamonds)
and after (crosses) renormalization (N=128).}
 \label{fig2_spectrum_wk}
\end{figure}

First, for the FPU system in thermal equilibrium \cite{check_eq},
with moderate and strong nonlinearities $\beta=1, 8, $ and $32$,
we measured the power spectrum $\langle|a_k|^2\rangle$ as a
function of $\omega_k$, where
\textbf{$a_k=(P_k-i{\omega}_{k}Q_k)/{\sqrt{2{\omega}_{k}}}$} and
$\langle\cdot\rangle$ denotes the time averaging. Although it was
expected that, for weak
nonlinearity, $\langle|a_k|^2\rangle=T/\omega_k$, where $T$ is an
effective temperature (Rayleigh-Jeans distribution for
waves~\cite{ZLF}), it is surprising to find that the same scaling
holds even for strong nonlinearities as shown in Fig.~\ref{fig1_ak}.
To understand why $\langle|a_k|^2\rangle$ scales as $\omega_k^{-1}$ even in the strongly
nonlinear limit, we
computed the $\omega$-$k$ spectrum of $a_k(t)$ as shown in
Fig.~\ref{fig2_spectrum_wk}(a).  In the weakly nonlinear limit the
spectrum would have resonant peaks along a curve given by the linear
dispersion relation $\omega_k=2\sin(k\pi/N)$. We observe that, when the
nonlinearity is no longer small, the
resonances move to higher frequencies, as the dispersion relation is
renormalized by the nonlinear part $V(Q)$. The renormalized dispersion
relations is
indicated by the sine-like structure of the resonance peaks (the
solid line in Fig.~\ref{fig2_spectrum_wk}(a)).  We verified
numerically that the main contribution of the nonlinear potential
energy ( $\sim 80$\%) comes from the terms $Q_kQ_lQ_m^*Q_s^*$
constrained on $k+l=m+s$. We note that when $k=m$ and $l=s$ or $k=s$
and $l=m$, these terms can be combined with the quadratic part of
(\ref{Ham_PQ}) to effectively renormalize the frequency $\omega_k$.
More specifically, Hamiltonian (\ref{Ham_PQ}) can be rewritten as
the sum of a new renormalized quadratic part and a remaining
nonlinear part:
$H=\widetilde{H}_{2}+\widetilde{H}_{4}$
where
$
\widetilde{H}_{2}=\frac{1}{2}\sum_{k=1}^{N-1}\left(|P_k|^2+\widetilde{\omega}_{k}^2|Q_k|^2\right)
$ with the renormalized dispersion relation,
\BEA \widetilde{\omega}_{k}=\eta\omega_k,~
\eta=\sqrt{1+\frac{3\beta}{2N}\sum_{l=1}^{N-1}\langle|Q_l(t)|^2\rangle\omega_l^2}.\label{wkr}
\EEA
Note that the frequency renormalization factor $\eta$ does not
depend on $k$.  Fig.~\ref{fig2_spectrum_wk}(b) shows how the
renormalized frequency $\widetilde\omega_k$ depends on $\beta$.  The
upper curve was produced from the numerical spectrum --- the abscissas
of the peaks of the sine-like curves were measured (e.g., from
Fig.~\ref{fig2_spectrum_wk}(a) for $\beta=1$, $\omega_{k=N/2}$=3.3
for $N=128$).  The curve with pluses was obtained using Eq.~(\ref{wkr}).
Both seem to possess a scaling $\sim\beta^{0.2}$ dependence (see
Fig.~\ref{fig2_spectrum_wk}(b)). The discrepancy between the
analytical and numerical curves can be attributed to the fact that
in the derivation of formula (\ref{wkr}) we took into account only
4-wave processes. In principle, we can take into account higher
order processes to obtain a more accurate analytical estimate of the
renormalizing factor.

To further study the renormalization of interactions we measured the
ratio of the quartic to the quadratic parts of the energy
\textit{before} and \textit{after} renormalization procedure (i.e.,
$H_4/H_2$ vs $\widetilde{H}_4/\widetilde{H}_2$) for different values
of $\beta$ with energy fixed (Fig.~\ref{fig2_spectrum_wk}(c)).
This figure also shows that the effective renormalized
linear part becomes more dominant. Therefore even for strongly
nonlinear regimes, with $\beta$ as large as 128, the
\textit{renormalized waves} (or purely nonlinear phonons) have weakened
interactions. Note that the resonance of $a_k(\omega)$ has a finite
width (shown with the dashed lines in Fig.~\ref{fig2_spectrum_wk}(a)
which are the level of $\int |a_k|^2(w)dw/\max_w|a_k|^2$ for each
$k$). We note that for $k=N/2$ (the highest mode in the system) the
resonances become the broadest. According to WT~\cite{ZLF}, the
energy exchange among waves occurs on the resonance manifold given
by $k_1+k_2=k_3+k_4$ and
$\omega_{k_1}+\omega_{k_2}=\omega_{k_3}+\omega_{k_4}$. Although one
can show that there are no exact resonances in $\beta$-FPU chains on
the discrete lattice, the nonlinearity induced near resonance interactions (i.e.
$|\omega_{k_1}+\omega_{k_2}-\omega_{k_3}-\omega_{k_4}|<\delta$ where
$\delta$ is a resonance width) can occur. This allows
us to use the WT theory if the interaction is considered to be weak.
In order to characterize the system as weakly nonlinear waves, the
renormalized waves are described by the new normal variables
$
\widetilde{a}_k=(P_k-i\widetilde{\omega}_{k}Q_k)/{\sqrt{2\widetilde{\omega}_{k}}}.
$
The relationship between bare $a_k$ and renormalized
$\widetilde{a}_k$ is $
a_k=\left[(\sqrt\eta+1/{\sqrt\eta})\ta_k+(\sqrt\eta-1/{\sqrt\eta})\ta_{-k}^*\right]/2.
$ Under the random phases approximation~\cite{Naz} for $a_k$ we have
$
\langle|\ta_k|^2\rangle=2\eta/(1+\eta^2)\langle|a_k|^2\rangle.
$ Therefore the power spectrum has scaling $\omega_k^{-1}$ for both
$a_k$ and $\ta_k$. (Note that $\langle|a_k|^2\rangle$ is shown in
Fig.~\ref{fig1_ak}).
  As a consequence, the effective
temperature, $\widetilde{T}$, for the renormalized waves, is related
to the bare temperature $T$ by
$\widetilde{T}=[2\eta^2/(1+\eta^2)]T$.

\begin{figure}
\vskip -0.3in \epsfxsize=2.7in \epsfysize=2.4in
 \epsffile{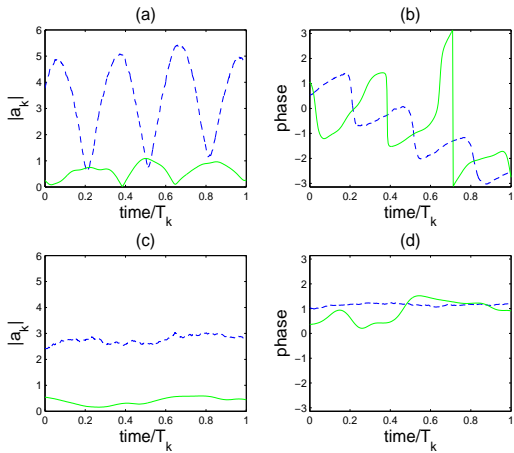}
 \vskip 0.1in
\caption[]{\footnotesize\baselineskip11pt (a) and (b) $|a_k|$ and
phase respectively, using the bare linear dispersion
$\omega_k=2\sin(\pi k/N)$. (c) and (d) $|\ta_k|$ and phase
respectively using the renormalized dispersion
$\widetilde{\omega}_k$ (Eq.\ref{wkr}). Dashed: $k=1$, solid: $k=20$.
The dispersion time scale is described by $T_k=2\pi/\omega_k$ for
(a) and (b) and $T_k=2\pi/\widetilde{\omega}_k$ for (c) and (d). (Note that according to our definition
$a_k=|a_k|e^{i(\phi+\omega_k t)}$, the phase does not include the linear
rotation.)}
\label{fig3_a}
\end{figure}

The renormalization picture is further corroborated by the time
evolution of each individual wave $\widetilde{a}_k$.  For the bare
waves $a_k$, there are large temporal modulations in both modulus
and phase as a result of strong nonlinear interaction among these
modes on the linear dispersion time scale (as shown in
Fig.~\ref{fig3_a}(a) and (b)). In weakly nonlinear systems, wave
amplitudes and corresponding phases are expected to evolve slowly on
the linear dispersion time scale. For renormalized waves
(Fig.~\ref{fig3_a}(c) and (d))these modes indeed have
characteristics of weakly interacting waves, with a small modulation
in $|\widetilde{a}_k|^2$ and phase.

\begin{figure}
\vskip -0.4in \epsfxsize=2.8in
 \epsffile{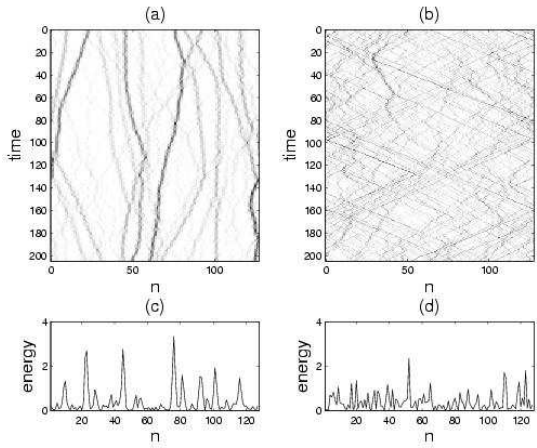}
\caption[]{\footnotesize\baselineskip11pt Energy density evolution
of
 the transient state (a) and (c), of the thermalized state (b) and
 (d), respectively ($\beta=1$ and $H=200$).  In (a) and (b) the darker
 strips correspond to high energy localizations. (c) and (d) are the
 snapshots of energy density.}\label{fig4_compare}
\end{figure}

Now we turn to the numerical evidence for the persistence of DBs in
the thermalized $\beta$-FPU chain.  As observed before~\cite{PhysD},
there are DBs present in the transient state.  As nonlinearity
increases, the duration of transient becomes shorter.  After the
energy redistributes among all the modes to achieve thermal
equilibration, our simulations show that the spatially localized,
high frequency excitations still exist. These DBs can interact with
each other and may be destroyed by collision processes with other
DBs or with the renormalized waves.  The spatial structure of these
excitations very much resembles the idealized breather oscillations
in the absence of spatially extended waves: they ``live'' above the
high frequency edge of the dispersion band and their lifetime is
sufficiently long (on the order of 10-100 DB oscillations) to behave
like a quasiparticle. Note that, under certain conditions,
supersonic solitons may arise from the $\beta$-FPU system as another
kind of localized excitations~\cite{Kink,Zhang}. However, they were not
observed in our thermalized system. Fig.~\ref{fig4_compare} is
the energy density plot which shows the time
evolution of energy of each particle for the transient (recording
starting time $T_0=5\times 10^2$) and thermalized ($T_0=5\times
10^5$) states, respectively. Fig.~\ref{fig4_compare}(c) and (d)
display the energy as a function of site at  $T_0$ corresponding to
Fig.~\ref{fig4_compare}(a) and (b) respectively.  In the transient
case (Fig.~\ref{fig4_compare}(a)) the spatially localized objects
(dark stripes) that carry sufficiently large amount of energy are
clearly observed. Fig.~\ref{fig4_compare}(c) is a snapshot of the
energy density plot (\ref{fig4_compare}(a)) at $T_0$. Here the DBs
are seen as localized peaks~\cite{PhysD}.  After thermalization the
spatial structure looks different (Fig.~\ref{fig4_compare}(b)).  The
system now consists of the renormalized waves (straight
cross-hatch traces in Fig.~\ref{fig4_compare}(b)). On the top of
these waves, the localized structures similar to DBs manifest
themselves as the wavy dark trajectories (in
Fig.~\ref{fig4_compare}(b)).  Although the snapshot
(Fig.~\ref{fig4_compare}(d)) of the energy density plot
(Fig.~\ref{fig4_compare}(b)) indicates that in thermal equilibrium
the energy is more evenly distributed among particles, spatially
localized structures are clearly observed.

\begin{figure}
\vskip -0.35in \epsfxsize=2.8in \epsfysize=2.8in
 \epsffile{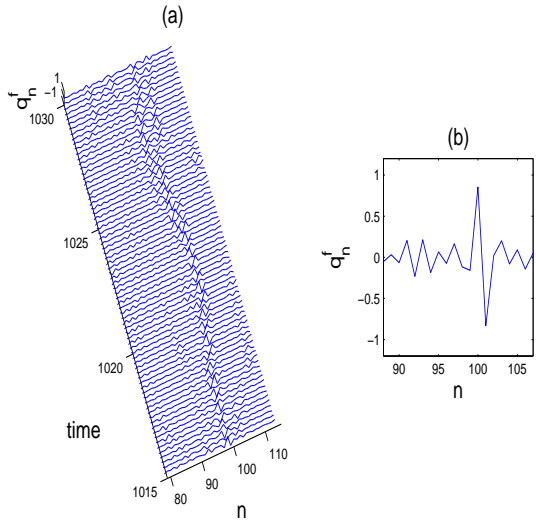}
 \vskip 0.1in
\caption[]{\footnotesize\baselineskip11pt (a) Evolution of a
discrete breather in thermal equilibrium. (b) typical snapshot of
the breather. $\beta=25$, $H=200$.}\label{fig5_space_time}
\end{figure}

\begin{figure}
\epsfxsize=2.8in \epsffile{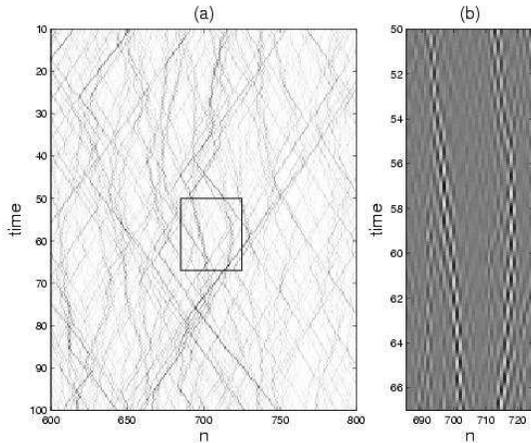}
\caption[]{\footnotesize\baselineskip11pt Turbulence of discrete
breathers (N=1024): (a) evolution of energy density, (b) zoomed in
$q_n^f(t)$ of the area indicated by the rectangle in (a). }
\label{fig7_eqp}
\end{figure}

Since there are renormalized waves in the system, which also carry
energy, we need to find a way to distinguish between these waves and
DBs. We use a frequency filter that cuts out the lower side of the
Fourier spectrum and leaves the high frequency part unmodified,
i.e., $f(g(n,t))=\Re F^{-1}(H_{\omega}(F(g)))$, where $\Re$ denotes
the real part, $F$ is a time Fourier transform, $H_{\omega}$
eliminates all
frequencies below $\omega_{cut}$ and $g(n,t)$ is a dynamical variable
that
is being filtered. By applying this filter to the displacement $q_n$
to obtain $q_n^f\equiv f(q_n)$, we can show the existence of DBs
even for strong nonlinearities, for example, $\beta=25$.
Fig.~\ref{fig5_space_time}(a) shows a clear
example of a DB excitation reconstructed using the filtered
$q_n^f$ with $\omega_{cut}=7$.  Fig.~\ref{fig5_space_time}(b) shows
a typical spatial profile of the DB taken from
Fig.~\ref{fig5_space_time}(a), which strongly resembles the
idealized DB~\cite{Cai}. Finally, in Fig.~\ref{fig7_eqp}, we present the
evidence that there
is a \textit{turbulence} of DBs, which chaotically ride on renormalized
waves. The corresponding energy density
distribution along with the distribution of filtered displacement in
a zoomed region $q_n^f(t)$ is displayed in Fig.~\ref{fig7_eqp}(a)
and (b), respectively.  After the lower modes from the displacement
$q_n$ are filtered, one can clearly observe that the remaining high
frequency oscillations are spatially highly localized, with the same
characteristics as an idealized breather.  The detailed time dynamics
of the DB shows the main characteristics of breathers: the
values of $q_n^f$ change signs periodically (as indicated by the
alternating white and black spots along the trajectory) as the DB
moves in space, with a spatial span of 2 or 3
sites only, as seen in Fig.~\ref{fig7_eqp}(b).

In conclusions, we have presented an interesting
dynamical scenario of the $\beta$-FPU chains in thermal equilibrium:
 (i) For strong
nonlinearity the linear dispersion relation is effectively
renormalized, which allows one to treat even strongly nonlinear
systems as if
they were weakly nonlinear; (ii) On top of renormalized waves, the strongly nonlinear
system is also characterized by the turbulence of discrete
breathers.

{\bf Acknowledgments} we thank Sergei Nazarenko for discussions.

\vskip -0.1in

\end{multicols}

\begin{thebibliography}{2000}
\bibitem{NumInt} We have verified that (i) the choice of various
integrators, (ii) initial conditions of spatially more regular structures
than random data, do not affect the reported results.
We used the 6-th order Yoshida algorithm~\cite{Yosh},
a 2-nd order symplectic and adaptive
RK integrators. We ensured a high degree of
accuracy, e.g., the energy is conserved up to the 9-th digit for a
runtime $T=10^6$.
\end{thebibliography}
\end{document}